\documentclass[prb,groupedaddress,superscriptaddress,twocolumn,showpacs]{revtex4-1} 

\usepackage[T1]{fontenc}
\usepackage[utf8]{inputenc}
\usepackage{lmodern}
\usepackage{graphicx}
\usepackage{amsmath}
\usepackage{amssymb}
\usepackage{amsfonts}
\usepackage{xcolor}
\usepackage[colorlinks=true, linkcolor=black, hyperindex=true]{hyperref}
\usepackage{epstopdf}

\begin{document}
\title{Exciton binding energies and luminescence of phosphorene under pressure}
\author{L. Seixas}
  \email{seixasle@gmail.com}
	\affiliation{Centre for Advanced 2D Materials and Graphene Research Centre, National University of Singapore, Singapore 117542, Singapore}

\author{A. S. Rodin}
  \affiliation{Centre for Advanced 2D Materials and Graphene Research Centre, National University of Singapore, Singapore 117542, Singapore}

\author{A. Carvalho}
	\affiliation{Centre for Advanced 2D Materials and Graphene Research Centre, National University of Singapore, Singapore 117542, Singapore}

\author{A. H. Castro Neto}
	\affiliation{Centre for Advanced 2D Materials and Graphene Research Centre, National University of Singapore, Singapore 117542, Singapore}
	\affiliation{Boston University, 590 Commonwealth Avenue, Boston, Massachusetts 02215, USA}

\date{\today}


\begin{abstract}
The optical response of phosphorene can be gradually changed by application of moderate uniaxial compression, as the material undergoes the transition into an indirect gap semiconductor and eventually into a semimetal. Strain tunes not only the gap between the valence band and conduction band local extrema, but also the effective masses, and in consequence, the exciton anisotropy and binding strength. In this article, we consider from a theoretical point of view how the exciton stability and the resulting luminescence energy evolves under uniaxial strain. We find that the exciton binding energy can be as large as 0.87 eV in vacuum for 5~\% transverse strain, placing it amongst the highest for 2D materials. Further, the large shift of the luminescence peak and its linear dependence on strain suggest that it can be used to probe directly the strain state of single-layers.
\end{abstract}

\maketitle

\section{Introduction}

The discovery of graphene ten years ago\cite{novoselov2004electric,neto2009electronic} triggered findings of a plethora of novel two-dimensional materials with unprecedented physical phenomena and potential technological applications\cite{schwierz2010graphene}. The wide variety of these novel 2D materials include insulators like hexagonal boron nitride (h-BN)\cite{alem2009atomically} and graphane\cite{elias2009control}, superconductors like niobium diselenide (NbSe$_{2}$)\cite{novoselov2005two}, and semiconductors like molybdenum disulfide (MoS$_{2}$) as well as other transition metal dichalcogenides (TMDC)\cite{novoselov2005two,mak2010atomically}.

\begin{figure}[!htb]
    \centering
        \includegraphics[width=0.49\textwidth]{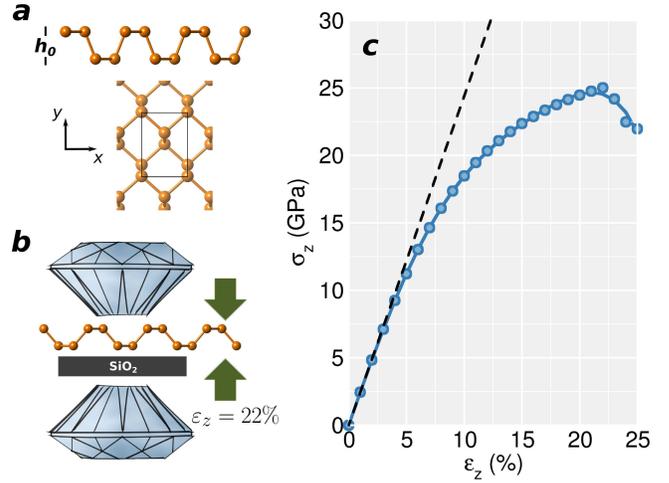}
    \caption{(color online) \textbf{Elastic properties of strained phosphorene.} (a) Ball-and-stick representation of unstrained phosphorene from sideview (top) and topview (bottom). (b) Schematic representation of the strained phosphorene in a diamond anvil cell. The green arrows indicate the direction in which the force was applied. (c) \textit{Ab initio} stress-strain curve as function of the uniaxial strain $\varepsilon_z$. Polynomial regression is shown with solid blue line, and dashed tangent line at $\varepsilon_z = 0$}
    \label{fig:Fig1}
\end{figure}

Recently, the paradigm to obtain novel two-dimensional materials from the exfoliation of layered crystals gave rise to the single-layer black phosphorus, also known as phosphorene \cite{rodin2014strain,liu2014phosphorene}. Phosphorene is a two-dimensional semiconductor formed exclusively by phosphorus atoms in a puckered anisotropic rectangular lattice due to the $sp^3$ hybridization with lone pairs [See Fig. \ref{fig:Fig1}(a)]. The single-layer and few-layer phosphorene can be obtained from mechanical exfoliation\cite{liu2014phosphorene} or plasma-assisted exfoliation of 3D layered crystal of black phosphorus (BP)\cite{lu2014plasma} weakly bound by van der Waals interaction. Since phosphorene is a homopolar semiconductor, it shows some advantages when compared with TMDC, such as electronic inactivity of grain boundaries and native point defects\cite{liu2014two}. Futhermore, the range of bandgaps from 0.3 eV (BP) to 1.6 eV (single-layer)\cite{PhysRevB.89.235319} and high hole mobility allow the production of field-effect transistors with on/off ratio up to $10^5$ at room temperature\cite{li2014black,liu2014phosphorene,qiao2014high}. The puckered anisotropic geometry of phosphorene also allows bandgap engineering with in-plane\cite{PhysRevB.90.085402,elahi2014modulation,wei2014superior} and out-of-plane strain\cite{rodin2014strain,jiang2014negative}, resulting in direct-indirect bandgap and semiconductor-metal transitions\cite{PhysRevB.90.085402,rodin2014strain}. 

One of the extraordinary phenomena observed in 2D semiconducting materials is the high stability of excitons, both due to the confinement of electron and hole and to the reduced screening of the Coulomb interaction. Concordingly, excitons dominate the optical spectra of 2D semiconductors, and can even be detected at room temperature.\cite{ye-nature-513-214,chernikov-PRL-113-076802} Measured exciton binding energies (EBEs) on monolayer transition metal dichalcogenides on SiO$_2$ are about 0.3-0.4 eV for sulfides and 0.6 eV for selenides,\cite{ye-nature-513-214,chernikov-PRL-113-076802,he-PRL-113-026803,wang-1404.0056,stroucken-1404.4238} and in vacuum are expected to reach values as high as 0.7~eV in WS$_2$\cite{ye-nature-513-214} and 1.1 eV in MoS$_2$.\cite{komsa-PRB-86-241201} These values are much higher than in bulk ($\sim$ 0.1 eV for WS$_2$\cite{beal-jphysc-5-3540}) and three dimensional semiconductors, where excitonic effects can be negleted from optical spectra to a good approximation. In suspendend (freestanding) phosphorene, the EBE has also been predicted to be 0.8 eV,\cite{tran-PRB-89-235319} even though phosphorene has a much smaller quasi-particle bandgap than WS$_2$ or MoS$_2$.\cite{tran-PRB-89-235319,ye-nature-513-214,komsa-PRB-86-241201} Also recently, the phosphorene EBE was calculated with in-plane strain in elastic regime \cite{peeters-2014}.

In this article, we consider the electronic and optical properties of phosphorene with uniaxial out-of-plane strain, and show that under such conditions the exciton binding energy can still be further increased. Moreover, we show that in an extended range of applied stress (both in the elastic and plastic regime), and specially near the point when the bandgap vanishes, the exciton binding energies are comparable to the quasi-particle bandgap. Thus, it is necessary to take into account the strength of the exciton binding when interpreting the evolution of the luminescence under stress.

The strain regime investigated here is comparable to what would be achieved in ideal conditions using a diamond anvil cell (DAC), as shown in Fig. \ref{fig:Fig1}(b). The compressive force is applied perpendicular to the phosphorene plane, and uniformized by a pressure-transmitting medium. We adopt the assumption that for a 2D material, the strain imposed by such a device setup can be considered uniaxial. This approximation is based on the small lateral cross section of phosphorene, which is orders of magnitude smaller than the area of mechanically exfoliated flakes. Due to phosphorene's high flexibility, strain as high as 22\% can be reached in the DAC, thus entering into the plastic regime of the material. 

\section{Methods}

The exciton binding energies for the 2D material with static dielectric constant $\epsilon$ were calculated with the dielectric screening induced by surrounding materials with dielectric constants $\epsilon_1$ (above) and $\epsilon_2$ (below). The screening effect of these dielectric materials on the phosphorene can be measured by mean dielectric constant $\kappa = \frac{\epsilon_1 + \epsilon_2}{2}$. The effective interaction between electrons and holes is given by the Hamiltonian
\begin{equation}
  H = \frac{p^2_x}{2\mu^{zz}} + \frac{p^2_y}{2\mu^{ac}} + V_{2D}(r),
\end{equation}
where $\mu^{zz}$ and $\mu^{ac}$ are the reduced effective masses (reciprocal mean of effective electron and hole masses) in \textit{zigzag} and \textit{armchair} directions, and $V_{2D}$ is the Keldysh potential given by
\begin{equation} \label{eq:keldysh}
  V_{2D}(r) = \frac{\pi e^2}{2\kappa r_0} \left[ H_0\left(\frac{r}{r_0}\right) - Y_0\left(\frac{r}{r_0}\right) \right],
\end{equation}
where $H_0$ and $Y_0$ are the Struve function and Bessel function of the second-kind, $r_0 = \frac{2\pi\zeta^{2D} }{\kappa}$, and $\zeta^{2D}$ is the 2D electric susceptibility. The EBE was calculated applying the Numerov method to solve the problem with the screening Keldysh potential \eqref{eq:keldysh}, as described in Ref. \onlinecite{rodin2014strain}. 

First-principles calculations based upon density functional theory (DFT)\cite{PhysRev.136.B864,PhysRev.140.A1133} were performed as implemented in the \textsc{PWscf} code of the \textsc{Quantum Espresso} package\cite{giannozzi2009quantum}. Norm-conserved pseudisation of external potential were performed with the Troullier--Martins parameterization\cite{troullier1991efficient}. We used energy cutoff of 70 Ry for the Kohn--Sham orbitals and $k$-points grid of $40 \times 40 \times 1$ in Monkhorst--Pack algorithm\cite{PhysRevB.13.5188} for phosphorene single-layer and $40 \times 40 \times 15$ for the weakly interacting black phosphorus\cite{PhysRevB.90.075429}. For the exchange-correlation functional we used the Perdew--Burke--Ernzerhof (PBE) approximation\cite{perdew1996generalized}. The strained phosphorene were completely relaxed until reaching the convergence criteria of residual forces smaller than $25$ meV/\AA\ and residual total energies smaller than $10^{-6}$ Ry. For the dielectric tensor calculation, we applied a scissor operator of 0.72 eV for the Kohn--Sham eigenvalues obtained from the comparison of the DFT bandgap from the GW calculations\cite{PhysRevB.90.075429}. Hybrid functional calculations based upon HSE06\cite{heyd2003hybrid,heyd2006erratum} functional were performed with PBE relaxed geometries. The small $q$-vectors divergence to the Coulomb potential was treated in the Gygi--Baldereschi approach\cite{gygi1986self}, and the three dimensional $q$-vector mesh $4\times4 \times 1$ was used for the Fock operator.

\section{Results}

In order to understand the uniaxial out-of-plane strain effect on phosphorene, we studied the phosphorene electronic properties by varying the layer height $h$, i.e. the distance between phosphorus planes on both sides of the same layer. The atomic positions and cell vectors were relaxed in plane under this fixed height constraint. The resulting stress-strain curve is shown in Fig. \ref{fig:Fig1}(c). 

The stress was calculated by Hellmann--Feynman theorem forces acting on the Born--Oppenheimer potential energy surfaces. The forces on the phosphorus atoms in a unit cell with area $A$ result in a stress $\sigma_z$ that can evolve non-linearly with the strain $\varepsilon_z = 1-\frac{h}{h_0}$, where $h_0$ is the phosphorene height in vacuum (no pressure). As the strain increases, the lattice parameters $a$ and $b$ of the relaxed variable-cell change so that the area of the strained unit cell is greater than area $A_0 = $ 15.34\ \AA$^2$ (in vacuum).

The \textit{ab initio} stress data was fitted by polynomial for the stress-strain curve, shown in Fig. \ref{fig:Fig1}(c) by the solid blue line\cite{Note1}. The Young's modulus is obtained from the first-order derivative of the stress-strain curve at $\varepsilon_z = 0$, i.e.,
\begin{equation}
  Y=\left.\frac{\partial \sigma_z}{\partial \varepsilon_z}\right|_{\varepsilon_z=0} = 242\ \mathrm{GPa}.
\end{equation}

The phosphorene elastic regime is inferred from the tangent line with slope equal to the Young's modulus, shown by the dashed black line in the Fig. \ref{fig:Fig1}(c). For strains of 3--4~\% (proportionality limit), the stress-strain curve is approximately linear. In this range, we say that phosphorene is in the elastic regime. Above $\varepsilon_z = 4~\%$, the phosphorene is in the plastic regime. The stress-strain curve increases monotonically until a local maximum at $\varepsilon_z = 22~\%$. The maximum stress, also called Yield strength, is about 25 GPa.

\begin{figure*}[hbt]
    \centering
        \includegraphics[width=0.96\textwidth]{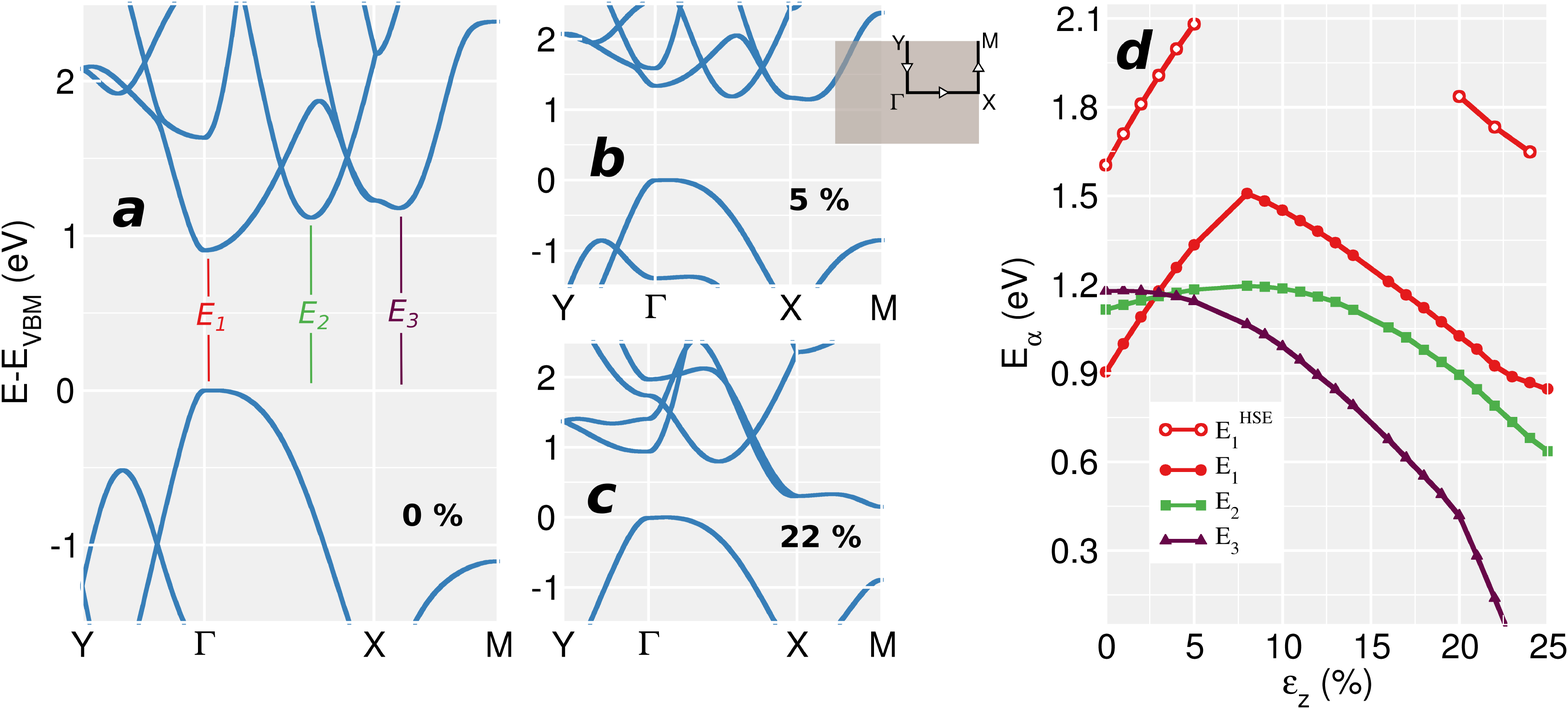}
    \caption{(color online) \textbf{Electronic properties of strained phosphorene.} Electronic band structure of strained phosphorene with: (a) $\varepsilon_z = 0.0$, (b)  $\varepsilon_z = 5.0$~\%, (c)  $\varepsilon_z = 22.0$~\%. (d) Direct ($E_1$) and indirect bandgaps ($E_2$ and $E_3$) as a function of strain $\varepsilon_z$. The band structures were obtained with PBE functional ($E_1$, $E_2$ and $E_3$) and HSE06 hybrid functional ($E_1^{HSE}$).}
    \label{fig:Fig2}
\end{figure*}

An alternative approach to the direct calculation of the stress can be conceived by spacial average of Nielsen--Martin\cite{PhysRevLett.50.697}. However, instead of normalizing using the supercell volume (with vacuum spacing), we normalized by the effective volume $V^{\rm eff}=Ah$, where $h$ is the phosphorene height under pressure and $A$ is the phosphorene unit cell area.

The Young's modulus for the $z$-direction (out-of-plane) is slightly larger than the average of the Young's moduli found in Ref. \onlinecite{wei2014superior} for the \textit{zigzag} and \textit{armchair} directions (in-plane strains). Despite this hardening in the $z$-direction, the low Young's modulus results in high flexibility and strongly tunable electronic properties with the uniaxial strain engineering.

To calculate the Poisson's ratio, we use the lattice constants $a$ and $b$ obtained from the variable cell relaxation dynamics with fixed strain $\varepsilon_z$. Using this method, we obtain the Poisson's ratio $\nu_{xz} = -0.06$ and $\nu_{yz} = 0.77$ at $\varepsilon_z = 0$. The auxetic property (negative Poisson's ratio) in the $x$-direction can be explained by the hinge-like geometry of the chemical bonds between the phosphorus atoms\cite{jiang2014negative}.

\begin{figure}[!htb]
    \centering
      \includegraphics[width=0.49\textwidth]{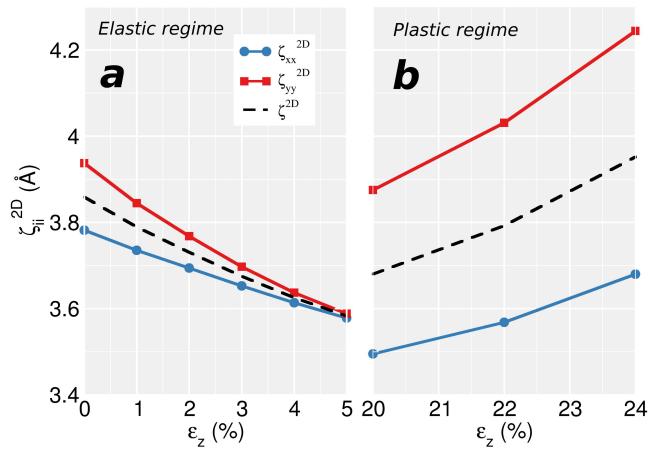}
    \caption{(color online) \textbf{2D electric susceptibility of strained phosphorene.} $\zeta^{2D}_{xx}$ (blue), $\zeta^{2D}_{yy}$ (red) and geometric mean $\zeta^{2D}$ (dashed black line) as a function of strain: (a) From 0.0 to 5.0~\%, (b) From 20.0~\% to 24.0~\%.}
    \label{fig:Fig3}
\end{figure}

\begin{figure}[!ht]
    \centering
        \includegraphics[width=0.49\textwidth]{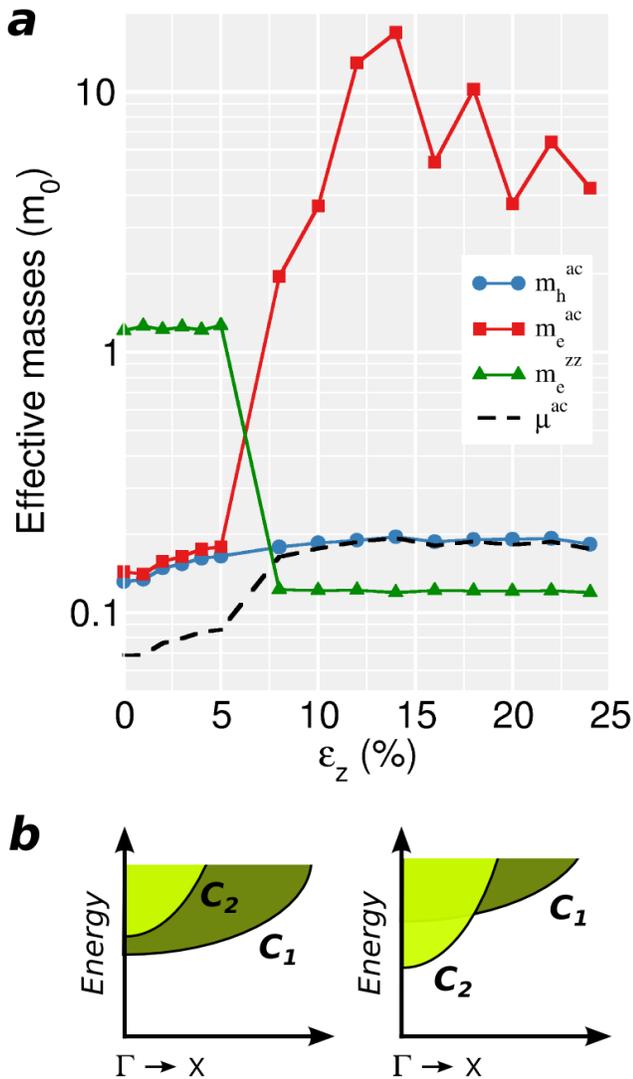}
    \caption{(color online) (a) Electrons and holes effective masses as a function of the strain in the \textit{armchair} and \textit{zigzag} directions, and reduced effective masses in \textit{armchair} direction. (b) Schematic representation of the band crossing between the conduction bands $C_1$ (dark green) and $C_2$ (light green) at $\varepsilon_z = 5$~\% (left) and $\varepsilon_z = 8$~\% (right).}
    \label{fig:Fig4}
\end{figure}

The band structures of phosphorene in equilibrium and under strain from 0 to 24~\% are shown in Fig. \ref{fig:Fig2}(a), (b) and (c). In equilibrium, phosphorene shows three valleys (local minima) in the conduction band, labeled $E_1$, $E_2$ and $E_3$. The evolution of these valleys with relation to the valence band maximum (VBM) is shown in Fig. \ref{fig:Fig2}(d). For strains smaller than $\varepsilon_z = 4$~\%, the band structure presents a direct bandgap at $\Gamma$-point. For strains close to $\varepsilon_z = 4$~\%, there a direct-indirect bandgap transition, with conduction band valley $E_2$ between $\Gamma$ and $X$ point. For strains larger than $\varepsilon_z = 5$~\% and smaller than $\varepsilon_z = 23$~\%, tha bandgap is indirect with minimum located between $X$ and $M$ point. From strains larger than $\varepsilon_z = 23$~\% there is a semiconductor-metal transition\cite{rodin2014strain}. Futhermore, for strains greater than $\varepsilon_z = 20$~\% the conduction band valley is located at $M$-point. This shift in the conduction band valley location is responsible for the change in the $E_3$ curve slope at 20~\% in Fig. \ref{fig:Fig2}(d).

\begin{figure*}[b!ht]
    \centering
        \includegraphics[width=0.96\textwidth]{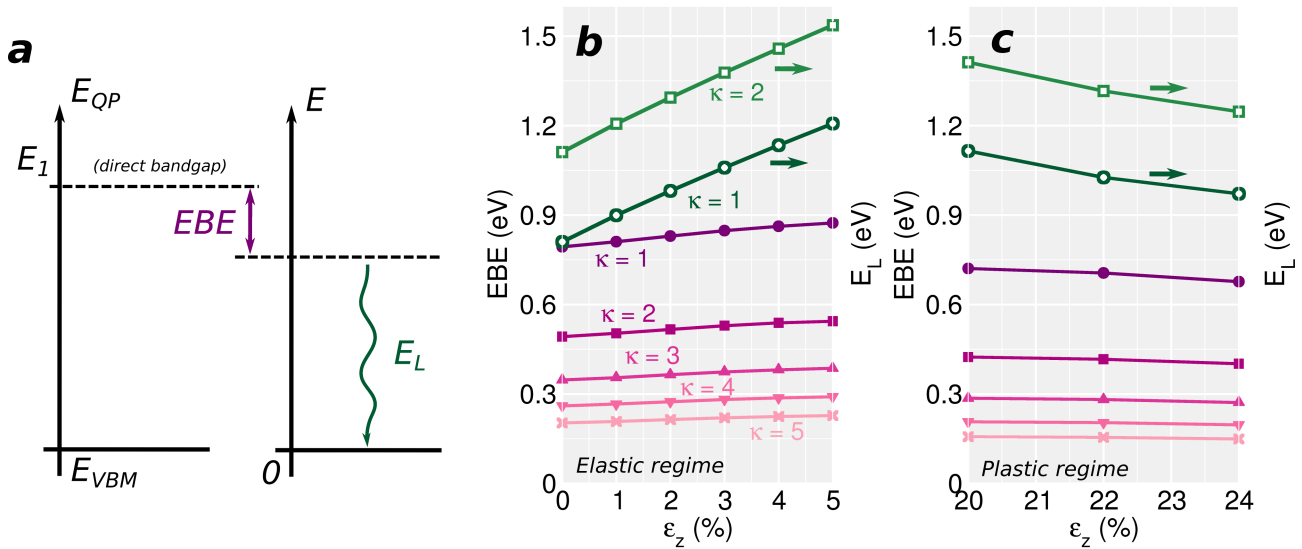}
    \caption{(color online) \textbf{Optical properties of strained phosphorene.} (a) Energy diagrams establishing a parallel between the quasi-particle gap in the one-electron formalism (left) and the exciton binding and photon energies in the many-body formalism (right). (b) and (c) Exciton binding energies ($EBE$) and Luminescenece energy ($E_L$) as function of strain $\varepsilon_z$.}
    \label{fig:Fig5}
\end{figure*}

We focus our attention on the $\Gamma$ excitons, that are expected to give rise to a luminescence peak throughout the whole range of strain, independently of the existence of a smaller indirect gap. The direct bandgap at $\Gamma$ takes lower values in two regions: (i) an approximately elastic region between 0 and 5~\%, and (ii) plastic region from 20~\% to 24~\%. The exciton binding energies were determined in this region, using as input the 2D electric susceptibility and the effective masses obtained using density functional theory.

The 2D electric susceptibility was calculated as described in Ref. \onlinecite{PhysRevB.90.075429,keldysh1979,berkelbach2013theory}. The 2D screening is characterized by the $\zeta^{2D}_{xx}$ and $\zeta^{2D}_{yy}$ parameters given by the variation of the electric permittivities $\epsilon_{xx}$ and $\epsilon_{yy}$ as a function of spacing between phosphorene layers. These 2D electric susceptibility are obtained from the fitting 
\begin{equation}
  \epsilon_{ii}(L) = 1 + \frac{4\pi \zeta^{2D}_{ii}}{L}, \ \ i=x,y,
\end{equation}
where $L$ is the unit cell height, ranging from $5.0$~\AA\ to $15.0$~\AA, as shown in Fig. \ref{fig:Fig3}. Note that the 2D electric susceptibilities $\zeta^{2D}_{xx}$ and $\zeta^{2D}_{yy}$ are more anisotropic in plastic regime than elastic regime. However, the exciton binding energy calculation depends only on the geometrical mean $\zeta^{2D} = \sqrt{\zeta^{2D}_{xx}\zeta^{2D}_{yy}}$.

The phosphorene anisotropy can also be seen from the large variation of the electrons' and holes' effective masses as a function of strain. Although the 2D electric susceptibility is averaged, the interaction between electrons and holes remains anisotropic through their effective masses in the \textit{zigzag} and \textit{armchair} directions, as shown in Fig. \ref{fig:Fig4}(a). While in the elastic regime the electrons' effective masses are light in the \textit{armchair} direction and heavy in the \textit{zigzag} direction, in plastic regime this is reversed. This reversion is explained by the band crossing between two conduction bands ($C_1$ and $C_2$), as shown in Fig. \ref{fig:Fig4}(b). This conduction band crossing occurs between $\varepsilon_z = 5$~\% and $\varepsilon_z = 8$~\%.

In the \textit{armchair} direction, for low strains, the holes' and electrons' effective masses are roughly equal, resulting in reduced effective masses ($\mu^{ac} \approx \frac{m_{e}^{ac}}{2} \approx \frac{m_{h}^{ac}}{2}$). However, for strains larger than 8~\%, the electrons' effective masses of $C_2$ state are so large that the reduced effective masses is approximately equal to holes' effective masses ($\mu^{ac} = m_{h}^{ac}$). In the \textit{zigzag} direction, the holes' effective masses are orders of magnitude greater than electrons' effective masses, so that we have $\mu^{zz} = m_{e}^{zz}$.

Based on the $\zeta^{2D}$ parameters and effective masses, we calculate the exciton binding energies (EBE) and luminescence energies ($E_{L}$) as a function of the strain $\varepsilon_z$, shown in Fig. \ref{fig:Fig5}. While the EBE increases 10~\% for $\varepsilon_z = 5$~\% (from 0.79 eV to 0.87 eV), the luminescence increases almost 50~\% for the same strain (from 0.81 eV to 1.21 eV). This variation of luminescence energies is explained by the $E^{HSE}_1$ conduction band valley behavior with the strain, that increase of 30~\% for $\varepsilon_z = 5$~\% (from 1.60 eV to 2.08 eV), as shown in Fig. \ref{fig:Fig2}(d). The EBE also depends strongly on the dielectric media into which phosphorene is immersed, parameterized by the mean permittivity $\kappa = \frac{\epsilon_1 + \epsilon_2}{2}$. The EBE and luminescence energies are shown in Fig. \ref{fig:Fig5}(b) in elastic regime and Fig. \ref{fig:Fig5}(c) in plastic regime.

\section{Discussion}

We have shown that the exciton binding energies are comparable in magnitude to the quasi-particle gaps and are sensitive to strain. Thus, the variation of the position of the luminescence peaks under strain in the ideal DAC experiment that inspired our work is affected by those two components. The resulting photon energy increases with strain in the elastic regime (up to 5~\%), in the plastic regime the trend is inverted and a redshift is observed.

In the elastic regime, the photon energy blueshifts by up to 0.4 eV for only  5~\% strain, and its variation is approximatelly linear, and nearly independent on the permittivity of the substrate. This suggests that the shift or broadening of the direct luminescence peak can be used as a direct way to probe the strain state of the material. This is an excellent alternative to Raman measurements, since the phosphorene Raman peaks are little changed by \textit{armchair} strain,\cite{fei2014lattice} and is also affected by the number of layers and the proximity to the edge.

Finally, we note that the exciton binding energy can be increased to 0.87 eV in vacuum for 
a modest strain of 5~\%. This value is in the range of exciton binding energies predicted for transition metal dicalcogenides, which however, have larger quasi-particle bandgaps.

\section*{Acknowledgements}
 L.S. acknowledges financial support provided by ``\textit{Conselho Nacional de Desenvolvimento Cient\'\i fico e Tecnol\'ogico}'' (CNPq/Brazil). The authors acknowledge the National Research Foundation, Prime Minister Office, Singapore, under its Medium Sized Centre Programme and CRP award ``\textit{Novel 2D materials with tailored properties: beyond graphene}'' (R-144-000-295-281). The first-principles calculations were carried out on the CA2DM and GRC high-performance computing facilities.


\end{document}